\documentclass[12pt,a4paper]{article}

\usepackage{amsmath,amsfonts,amssymb}
\usepackage[dvips]{graphicx}

\def\be{\begin{equation}}
\def\ee{\end{equation}}
\def\bea{\begin{eqnarray}}
\def\eea{\end{eqnarray}}

\begin{document}

\title{Simple dynamics on the brane.}

\author{
Marek Szyd{\l}owski \\
Astronomical Observatory, Jagiellonian University \\
Orla 171, 30-244 Krak{\'o}w, Poland\footnote{%
email: uoszydlo@cyf-kr.edu.pl} \\
\and
Mariusz P. D\c{a}browski \\
Institute of Physics, University of Szczecin \\
Wielkopolska 15, 70-451 Szczecin, Poland\footnote{%
email: mpdabfz@uoo.univ.szczecin.pl} \\
\and
Adam Krawiec \\
Institute of Public Affairs, Jagiellonian University \\
Rynek G{\l}{\'o}wny 8, 31-042 Krak{\'o}w, Poland\footnote{%
email: uukrawie@cyf-kr.edu.pl}}

\date{\today}

\maketitle

\begin{abstract}
We apply methods of dynamical systems to study the behaviour of
the Randall-Sundrum models. We determine evolutionary paths for all possible
initial conditions in a 2-dimensional phase space and we investigate the set of
accelerated models. The simplicity of our formulation in comparison to some earlier studies
is expressed in the following: our dynamical system is a 2-dimensional Hamiltonian
system, and what is more advantageous, it is free from the degeneracy of critical
points so that the system is structurally stable. The phase plane
analysis of Randall-Sundrum models with isotropic Friedmann geometry
clearly shows that qualitatively we deal with the same types of
evolution as in general relativity, although quantitatively there
are important differences.
\end{abstract}

PACS number(s): 04.20.Jb, 04.65.+e, 98.80.-k, 98.80.Hw

\section{Introduction}

The rapid development of particle-physics-motivated cosmology
represented mainly by superstring cosmology \cite{ven91,gv93,superjim} has
led to a change of views onto the standard problems of
inflationary cosmology such as the (past) horizon and flatness
problems \cite{guth,linde}. On the other hand, the astronomical observations of
supernovae Ia \cite{Supernovae} strongly suggest that the universe not only had possibly accelerated
at its early stages of evolution but it is also accelerating now.
This puts strong constraints onto the matter content of the
universe because only the exotic (negative pressure) matter in
standard cosmology can lead to acceleration. However, a
phenomenological nature of such an exotic matter in standard
models expressed in terms of the perfect fluid does not seem to
make enough connection with particle physics and that is why one
is looking for other, more physical, descriptions. Among them the
main proposal is quintessence or time-dependent scalar field
\cite{steinhardt} which substitutes an ordinary phenomenological
barotropic fluid. However, there are other interesting proposals
to express the phenomenon of acceleration which are related to
superstring or M-theory models. The best known are pre-big-bang
\cite{ven91,gv93}, brane \cite{hw,rs1,rs2} and
ekpyrotic models \cite{turok1,turok2}. In this paper we try to study the
standard cosmology problems (such as cosmic acceleration and past horizon
problems) within the framework of the brane universes. We do not
study, for instance, the future horizon problem following the
recent discussion inspired by the S-matrix formulation within the
superstring theories \cite{futurehor}. In the paper we are
interested mainly in the early stage of the evolution of the
universe although the supernovae data gives restrictions onto
this stage and they should be taken into account.

The idea of brane universes has been first presented by Ho\v{r}ava and
Witten \cite{hw} who considered strong coupling limit of heterotic $E_8 \times
E_8$ superstring theory, i.e., M-theory. This limit results
in `exotic' \cite{Visser85,Barcelo00} Kaluza-Klein type compactification of $N = 1$, $D =11$
supergravity on a $S^1/Z_2$ orbifold (a unit interval) in a similar
way as compactification of $N=1$, $D=11$ supergravity on a circle
$S^1$ results in strongly coupled limit of type IIA superstring
theory. In Ho\v{r}ava-Witten theory there exist two 10-dimensional
branes to which all the gauge interactions are confined, and they are
connected via the orbifold, with gravity propagating in all 11 dimensions.
After further compactification of Ho\v{r}ava-Witten models on a Calabi-Yau
manifold one gets an effective 5-dimensional theory which has been
applied to cosmology \cite{lukas,reall,jim,mpd00}.

Randall and Sundrum \cite{rs1,rs2} developed similar to Ho\v{r}ava-Witten
scenario which was mainly motivated by the hierarchy problem in
particle physics \cite{Arkani98,Arkani99}. As a result, they obtained a
5-dimensional spacetime (bulk) with $Z_2$ symmetry with two/one
3-brane(s) embedded in it to which all the gauge interactions are
confined. In one-brane scenario \cite{rs2} the brane appears at the $y=0$
position, where $y$ is an extra dimension coordinate and the 5-dimensional
spacetime is an anti-de Sitter space with negative 5-dimensional cosmological
constant. The extra dimension can be infinite due to the exponential `warp'
factor in the metric
\begin{equation}
\label{rsmet}
ds^2 = \exp{\left(-2\frac{\mid y \mid}{l}\right)} \left[ -dt^2
+ d\vec{x}^2 \right] + dy^2,
\end{equation}
and $l$ gives the curvature scale of the anti-de Sitter space.
In the simplest case the induced metric on a brane is a
Minkowski metric (energy momentum tensor of matter vanishes).
However, the requirement to allow matter energy-momentum tensor on
the brane leads to breaking of conformal flatness in the bulk, and
the metric (\ref{rsmet}) is no longer valid. This fact is
obviously related to the appearance of the Weyl curvature in the
bulk \cite{BDL,BDEL,roy01b}. The full set of 5-dimensional and projected
4-dimensional equations has been presented in
Refs.~\cite{Shiromizu00,Sasaki00,Mukhoyama00}. Global geometric
properties of such brane models have also been studied (see Ref.
\cite{Ishihara01}). Generalized bulk spacetimes different from those
of (\ref{rsmet}) have been found in Ref. \cite{roy01a}. Anisotropic Kasner
branes have been immersed in AdS bulk in Ref.~\cite{afrolov}.

Campos and Sopuerta \cite{Campos01a,Campos01b} used the dynamical system
methods to the analysis of the Friedmann, Bianchi I and V Randall-Sundrum brane
world type cosmological
models with a non-vanishing 5-dimensional Weyl tensor. They considered
the dynamics of this model in the form of a {\it
higher-than-two-dimensional}
dynamical system. Exact analytic brane configurations with a vanishing Weyl tensor with
perfect and viscous fluid have been presented in Refs. \cite{harko1,harko2}.
Coley \cite{Coley01a,Coley01b} also studied the dynamics of these Randall-Sundrum
models - he made a step towards Mixmaster (Bianchi IX) dynamics and found it was
not chaotic provided the matter had positive pressure. In this paper we show that
the dynamical system which
describes the evolution of the brane models (both isotropic (FRW) and
anisotropic (Bianchi I or Bianchi V) types)
can be represented in the simplest way in the form of a {\it two-dimensional\/}
Hamiltonian dynamical system. Such visualization has a great advantage
because we {\it avoid\/} the problem of degeneracy of critical points which
appeared in the {\it higher dimensional} phase space (see Ref.~[13] in
\cite{Campos01b}). It is well
known that the existence of such critical points is a possible reason
for the structural instability of a model. On the other hand, the representation of
dynamics as a one-dimensional Hamiltonian flow allows to make the
classification of possible evolution paths in the {\it configuration space} which
is complementary to phase diagrams. It also makes
simpler to discuss the physical content of the model. Finally, the
construction of the Hamiltonian allows to study quantum cosmology
on the brane as it was attempted in Refs. \cite{nunez1,nunez2}, in
full analogy to what is usually done in general relativity
\cite{DL95}.

In this paper we demonstrate the effectiveness of representation of dynamics
as a one-dimensional Hamiltonian flow. In this representation the phase
diagrams in a two-dimensional phase space allow to analyze the
{\it acceleration and (past) horizon} problems in a clear way. We reduce the dynamics
to a two-dimensional phase space with an autonomous system of
equations.

We deal with the full global dynamics of brane models whose
asymptotic states are represented by critical points of the
system. From theoretical point of view it is important how large
the class of accelerated models is. We will call this class of
accelerated models typical/generic, if the domain of acceleration in the
phase space is a non-zero measure. On the other hand, if only the
non-generic (zero-measure) trajectories are represented by
accelerated models, then the mechanism which drives these
trajectories should be called ineffective. Such a point of view is
a consequence of the fact that, if the acceleration is an attribute
of a trajectory which starts with a given initial conditions, it
should also be an attribute of a trajectory which starts with a
neighbouring initial conditions.

Our analysis of the brane-world type model may be considered as
complementary to Campos and Sopuerta's analysis \cite{Campos01a,Campos01b}
(see also \cite{Coley01a,Coley01b}). However, it is more advantageous in many
points. It is because our dynamical system is a 2-dimensional system which is the
smallest possible dimension to study isotropic cosmological systems
of equations. This allows to avoid huge redundancy of degenerated critical
points and trajectories which appear in a higher dimensional phase space (cf. phase diagrams of
\cite{Campos01a,Campos01b}) so to the structural instability.
Finally, for a two-dimensional Hamiltonian system one is able to
study all its properties in a configuration space rather than in
a phase space.

The paper is organized as follows. In Section 2 we present simple Hamiltonian
dynamics of the brane
universes. In Section 3 we discuss cosmological models in configuration space, stability
and the domains of the phase plane for
which cosmic acceleration and horizon problem avoidance appear. In Section 4
we present phase portraits for FRW models with dust ($\gamma =1$)
and domain-wall-like matter ($\gamma = 1/3$) \cite{Supernovae,stelbd93,Dabrowski96} which induces accelerated expansion (quintessence) in
standard general relativistic cosmology. In section 5 we present simple form of dynamical systems in which
constant coefficients play the role of the observational dimensionless density of matter
$\Omega$ parameters. In Section 6 we discuss our results.

\section{Simple dynamics of brane universes}

The 5-dimensional Einstein equations for the Randall-Sundrum (RS) model with
a brane at $y=0$ location are \cite{Shiromizu00,Sasaki00,Mukhoyama00}
\begin{equation}
\label{eq:1}
\tilde{G}_{\mu \nu}^{(5)} = \kappa_{(5)}^{2}
[-\Lambda_{(5)} g_{\mu \nu}^{(5)} + \delta(y)(-\lambda h_{\mu \nu}^{(4)}
+ T_{\mu \nu}^{(4)})]
\end{equation}
where $\tilde{G}_{\mu \nu}^{(5)}$ is a 5-dimensional Einstein
tensor, $g_{\mu \nu}^{(5)}$ is a 5-dimensional metric,
$\Lambda_{(5)}$ is a 5-dimensional cosmological constant, and
$h_{\mu \nu}^{(4)}=g_{\mu \nu}^{(5)}-n_{\mu}n_{\nu}$ is a 4-dimensional
induced metric, $n^{\alpha}$ a unit normal vector to the brane,
$T_{\mu \nu}^{(4)}$ -- a 4-dimensional energy-momentum tensor,
$\lambda$ is a brane tension, and
\begin{equation}
\label{eq:2}
\kappa_{(5)}^{2}=8\pi {}^{(5)}G_{N} = \frac{8\pi}{{}^{(5)}M_{p}^{3}}.
\end{equation}
where the 5-dimensional Planck mass ${}^{(5)}M_{p}$ is always much less than
the 4-dimensional ${}^{(4)}M_{p}$ as measured on the brane
${}^{(5)}M_{p} \ll {}^{(4)}M_{p}=1.2 \cdot 10^{19} \text{GeV}$.
This fact allows to solve the hierarchy problem in particle physics and
possibly reach the electroweak scale $\sim$ TeV for gravity in accelerators.

The induced 4-dimensional Einstein equations on the brane are
\cite{roy01b}
\begin{equation}
\label{eq:3}
G_{\mu \nu}^{(4)} = - \Lambda_{(4)} h_{\mu \nu}^{(4)} +
\kappa_{(4)}^{2} T_{\mu \nu} + \kappa_{(5)}^{2} \Pi_{\mu \nu} - E_{\mu \nu}
\end{equation}
where
\begin{gather}
\label{eq:4}
\kappa_{(4)}^{2} = 8 \pi {}^{(4)}G_{N} = \frac{8\pi}{{}^{(4)}M_{p}^{2}} =
\frac{\lambda}{6} \kappa_{(5)}^{4} \\
\Lambda_{(4)} = \frac{1}{2} \kappa_{(5)}^{2} \left[ \Lambda_{(5)} +
\label{eq:5}
\frac{1}{6} \kappa_{(5)}^{2} \lambda^{2} \right] \\
\label{eq:6}
\Pi_{\mu \nu} = \frac{1}{12} T T_{\mu \nu} - \frac{1}{4} T_{\mu \alpha}
T^{\alpha}_{\phantom{\alpha} \nu} + \frac{1}{24} g_{\mu \nu}
[3 T_{\alpha \beta} T^{\alpha \beta} - T^{2}],\\
T_{\mu\nu} = \rho V_{\mu}V_{\nu} + p h^{(4)}_{\mu\nu} ,
\end{gather}
where $V_{\mu}$ is the 4-velocity of an observer on the brane,
$\Lambda_{(4)}$ is the 4-dimensional cosmological constant on the brane
\cite{BDL,BDEL} and $E_{\mu\nu}$ is the correction which appears from the Weyl
tensor in the bulk which reads as \cite{roy01b,Coley01a,Coley01b}
\begin{equation}
E_{\mu\nu} = - \frac{6{\cal U}}{\lambda \kappa^2_{(4)}} \left[ V_{\mu}V_{\nu}
+ \frac{1}{3} h^{(4)}_{\mu\nu} + {\cal P}_{\mu\nu} + {\cal Q}_{\mu}V_{\nu}
+ {\cal Q}_{\nu}V_{\mu}\right] .
\end{equation}
Here ${\cal U}$ is an effective nonlocal energy density on
the brane which arises from the gravitational field in the bulk
which is not necessarily positive and reads
\begin{equation}
{\cal U} = -  \frac{1}{6} \kappa^2_{(4)} \lambda  E_{\mu\nu} V^{\mu} V^{\nu}.
\end{equation}
Since $E_{\mu\nu}$ is traceless, then its effective local pressure is $p = (1/3){\cal U}$.
On the other hand, an effective nonlocal anisotropic stress is
\begin{equation}
{\cal P}_{\mu\nu} = - \frac{1}{6} \kappa^2_{(4)} \lambda E_{[\mu\nu]},
\end{equation}
while an effective energy flux on the brane is
\begin{equation}
{\cal Q}_{\mu} = - \frac{1}{6} \kappa_{(4)} \lambda
\left( E_{\mu\nu}V^{\nu} + E_{\nu\mu}V^{\mu} \right)   .
\end{equation}

After admission the perfect fluid with barotropic equation of state
$p=(\gamma -1)\rho$, $\gamma \in [0,2]$, where $p$ the pressure and
$\varrho$ the energy density from (\ref{eq:6}) one has
\begin{equation}
\Pi_{\mu\nu} = \frac{1}{12} \varrho^2 V_{\mu}V_{\nu} +
\frac{1}{12} \varrho^2 (2 \gamma - 1) h^{(4)}_{\mu\nu}
\end{equation}
and the dynamics of homogeneous models
in the RS brane-world scenario can be described by the following set of
equations
\begin{align}
\label{eq:7}
\dot{H} &= - H^2 - \frac{(3\gamma - 2)\kappa_{(4)}^{2}}{6} \rho
- \frac{(3\gamma - 1)\kappa_{(4)}^{2}}{6\lambda} \rho^{2}
- \frac{2}{3}\sigma^{2} + \frac{\Lambda_{(4)}}{3}
- \frac{2}{\lambda \kappa_{(4)}^{2}} {\cal U} ,\\
\label{eq:8}
\dot{\rho} &= - 3\gamma H \rho , \\
\label{eq:9}
\dot{\cal U} &= - 4 H {\cal U} ,
\end{align}
where $H=d(\ln a)/dt$ is the Hubble function, $t$ is cosmological time,
the brane tension is
$\lambda >0$ (this condition allows to recover conventional gravity)
and $\kappa_{(4)} = 8\pi G_{(4)}$ is the
gravitational coupling constant ($c=1$). The function ${\cal U}$ enters
as a contribution from the Weyl tensor in the bulk.
The shear scalar $\sigma^2 = \frac{1}{2}
\sigma^{ab} \sigma_{ab}$ vanishes for the FRW models, whereas
for the anisotropic Bianchi type models
\begin{equation}
\label{eq:10}
\frac{d \sigma^{2}}{dt} = -6 H \sigma^{2} \quad \Leftrightarrow
\sigma^{2} = \sigma_{0}^{2} a^{-6}
\end{equation}
The first integral of the system (\ref{eq:7})--(\ref{eq:9}) is
the Friedmann equation \cite{Campos01b} which reads as
\begin{equation}
\label{eq:11}
H^{2} = \frac{\kappa_{(4)}^{2}}{3} \rho + \frac{\kappa_{(4)}^{2}}{6\lambda}
\rho^{2} - \frac{k}{a^{2}} + \frac{\sigma^{2}}{6} + \frac{\Lambda_{(4)}}{3}
+ \frac{2{\cal U}}{\lambda \kappa_{(4)}^{2}} ,
\end{equation}
where $k \in \{ 0, \pm 1\}$ -- the curvature index. One can easily see from
(\ref{eq:11}) that in the limit $\lambda \to \infty$ one recovers general relativity.
The main difference from the standard general relativistic Friedmann
equation is the appearance of the $\varrho^2$ correction. This
term comes as a contribution from the brane. However, in the
general case within the framework of AdS/CFT correspondence
\cite{maldacena} this correction may come from higher-derivative
terms in the curvature and higher-dimension operators connecting
the on-brane matter directly to the CFT \cite{march-russell}.

There are two invariant submanifolds: ${\cal U}=0$ and
$\sigma=0$. The system on the submanifold ${\cal U}=0$ corresponds to dynamics
with vanishing electromagnetic part of the Weyl tensor, whereas
the system on the submanifold $\sigma=0$ corresponds to the case
without anisotropy (FRW models). The intersection of these submanifolds
is also an invariant.

It is well-known that the first integral of
the FRW equation can be used to construct a Hamiltonian function.
We take advantage of this feature in the considered model.

\begin{figure}
\includegraphics[width=\textwidth]{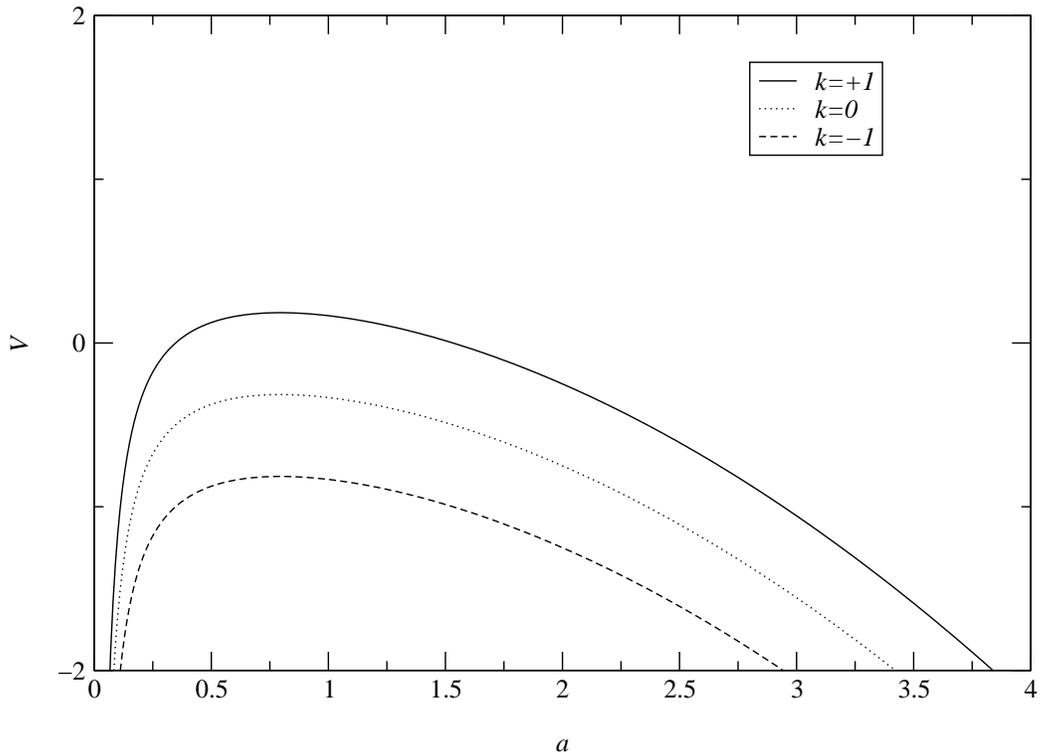}
\caption{The potential function $V(a)$ given by (\ref{eq:15})
for some trajectories depending on
the initial conditions in the phase plane for $k=0,-1,+1$
and $\gamma=1$ (dust). The horizontal line set by loci formed from
maxima of all potential functions separates the regions with
{\it deceleration} (left) and {\it acceleration} (right) (cf. Eq. (\ref{eq:14})).
Only the region with $V(a)<0$ has a physical sense.}
\label{fig:1}
\end{figure}

The integration of (\ref{eq:8})--(\ref{eq:9}) gives
\begin{equation}
\label{eq:12}
\rho = \rho_{0} a^{-3\gamma}, \qquad {\cal U} = {\cal U}_{0} a^{-4}.
\end{equation}
Therefore the right-hand side of the Rayuchaudhuri equation (\ref{eq:7}) can be expressed in
terms of the scale factor $a(t)$ as
\begin{multline}
\label{eq:13}
\ddot{a} = \left[ - \frac{(3\gamma -2)\kappa_{(4)}^{2} \rho_{0}}{6}
a^{-3\gamma} - \frac{(3\gamma -1)\kappa_{(4)}^{2} \rho_{0}^{2}}{6\lambda}
a^{-6\gamma} - \frac{2}{3} \sigma_{0} a^{-6} \right. \\
\left. + \frac{\Lambda_{(4)}}{3}
- \frac{2 {\cal U}_{0}}{\lambda \kappa_{(4)}^{2}} a^{-4} \right] a.
\end{multline}
The equation (\ref{eq:13}) can be rewritten in the form analogous to the
Newton equation of motion in the 1-dimensional configuration space
$\{ a\!: a \in \mathbb{R}_{+}\}$
\begin{equation}
\label{eq:14}
\ddot{a} = - \frac{\partial V}{\partial a} ,
\end{equation}
where the potential function
\begin{equation}
\label{eq:15}
V(a) = -\frac{\kappa_{(4)}^{2} \rho_{0}}{6} a^{-3\gamma +2}
- \frac{\kappa_{(4)}^{2} \rho_{0}^{2}}{12\lambda} a^{-6\gamma +2}
- \frac{\sigma_{0}^2}{6} a^{-4} - \frac{\Lambda_{(4)}}{6} a^{2}
- \frac{{\cal U}_{0}}{\lambda \kappa_{(4)}^{2}} a^{-2} + V_{0},
\end{equation}
and $V_{0} = \text{const}$. The plot of (\ref{eq:15}) for $\gamma = 1$
is given in Fig.\ref{fig:1}. The equation (\ref{eq:14}) has a simple
interpretation: the Randall-Sundrum models {\it accelerate}
wherever the potential $V(a)$ is a decreasing function of $a$
while they {\it decelerate} whenever the potential $V(a)$ is an
increasing function of $a$. Another important
conclusion from (\ref{eq:15}) is that for $\gamma > 1 (p > 0)$ $\rho^2$
contribution dominates over shear anisotropy near singularity so
the singularity in Bianchi models (especially in Bianchi IX) is
isotropic and that there is no chaos \cite{Coley01b}. In general relativity it
is possible only for a meaningless fluid with $\gamma > 2$ $(p >
\rho)$. The first integral of (\ref{eq:14}) is
\begin{equation}
\label{eq:16}
V(a) + \frac{\dot{a}^{2}}{2} = V_{0} - \frac{k}{2}.
\end{equation}
Now we construct the {\it Hamiltonian function}
\begin{equation}
\label{eq:17}
\mathcal{H} \equiv \frac{\dot{a}^{2}}{2} + V(a),
\end{equation}
and then trajectories of the system lie on the energy level
$\mathcal{H} \equiv E = \text{const}$. The advantage of having the
Hamiltonian function given by Eq. (\ref{eq:17}) is that one is then
easily able to canonically quantize the system and allow quantum
cosmological framework on the brane \cite{nunez1,nunez2} in full analogy to general relativity \cite{DL95}.
Finally, we obtain the dynamics reduced to the Hamiltonian flow in the 1-dimensional
configuration space
\begin{equation}
\label{eq:18}
\mathcal{H} = \frac{\dot{a}^{2}}{2} + V(a) = 0 \;,
\end{equation}
with $V(a)$ given by (\ref{eq:15}) and $V_0 = k/2$.

Now the physical trajectories lie on {\it the zero-energy level},
$\mathcal{H} = E = 0$, which coincides with the form of the first integral.
\begin{figure}
\includegraphics[width=\textwidth]{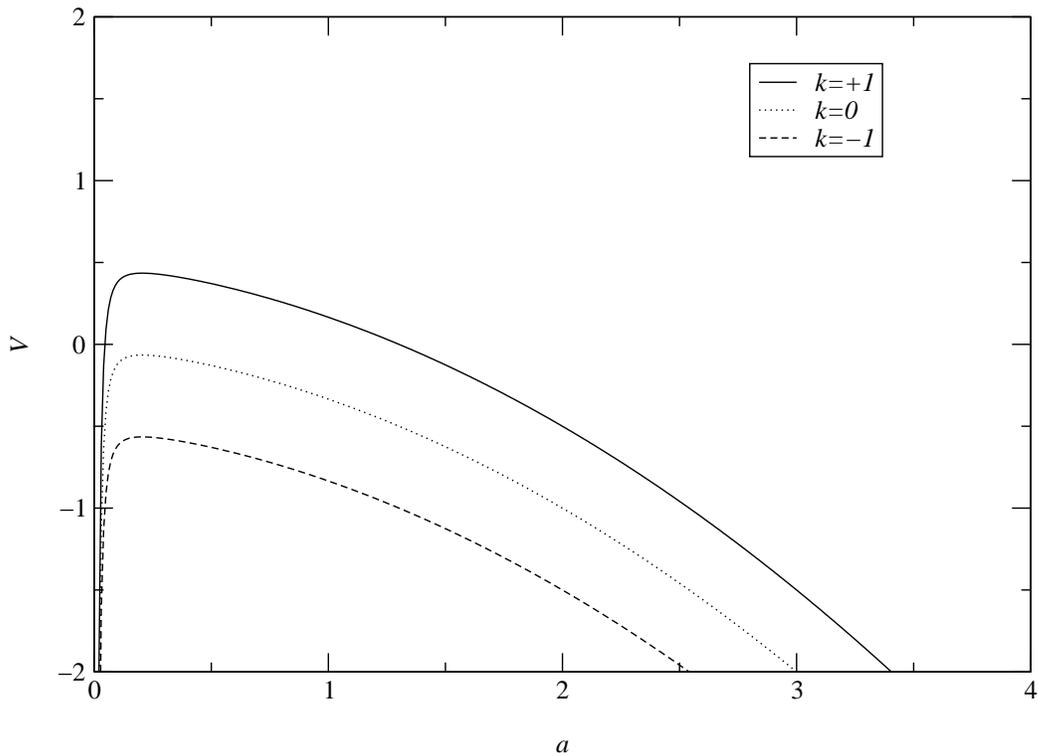}
\caption{The potential function $V(a)$ given by (\ref{eq:15})
for some trajectories depending on
the initial conditions on the phase plane for $k=0,-1,+1$
and $\gamma=1/3$ (domain walls) which induces accelerated expansion in standard general
relativistic cosmology. Similarly as in Fig.~\ref{fig:1} there is the
horizontal line of potential maximum points which separates
to {\it deceleration} and {\it acceleration} regions (cf. Eq. (\ref{eq:14})).
However it appears for lower values of $a$.
}
\label{fig:2}
\end{figure}

The formalism can simply be generalized
to the case of a general form of the equation of state. We use this fact
and consider the equation of state for the (non-interacting) mixture
of dust-like matter ($p=0$) and an unknown component, labelled $X$,
with negative pressure. It is assumed that dominant $X$-component is
a perfect fluid with the equation of state specified by $p_{X} =
(w_{X}-1) \rho_{X}$ and $0 < w_{X} < 2/3$, which enables that component
to induce accelerated expansion in standard general relativistic cosmology
\cite{stelbd93,Dabrowski96}. The positive cosmological
constant $\Lambda_{(4)}$ corresponds to $w_{X}=0$. The supernovae observations
indicate that the dimensionless energy density of X-component $\Omega_{X0} \approx 0.7$ is large compared to dust-like
matter ($\rho_{X0} \ll M_{pl}^{4}$ --- natural order of magnitude of
the vacuum energy) \cite{Supernovae}.
The best-fit component is with $w_{X} \approx 1/3$ (see paper 4 in \cite{Supernovae}).
One often represents the
$X$-component in terms of a scalar field $\Phi$ (e.g., inflaton) with a suitable
potential (quintessence) \cite{maeda,diaz}.
Conservation of energy gives $\rho_{\Phi} \propto a^{m}$, $m=-3w_{\Phi}$,
$p_{\Phi} = (w_{\Phi}-1) \rho_{\Phi}$, $w_{\Phi} = \text{const}$.

The most general case of the equation of state is
\begin{equation}
\label{eq:a1}
p=p_{X} + 0 = \left[\gamma(a) - 1\right] \rho
= \frac{w_{X}}{\frac{\rho_{m0}}{\rho_{X0}} a^{3(w_{X}-1)} + 1} \rho
,
\end{equation}
with $w_{X} =$ const. and zero stands for the pressure of the dust.

Therefore the total energy $\rho = \rho_{m} + \rho_{X}$ changes as
\begin{equation}
\label{eq:a2}
\rho = \rho_{0} a^{-3} \exp\left( - \int_{0}^{a} \frac{3\gamma(a)}{a}
da \right).
\end{equation}
After substitution (\ref{eq:a1}) into (\ref{eq:a2}) we obtain
\begin{equation}
\label{eq:a3}
\rho = \rho_{X0} a^{-3w_{X}} \left[ \frac{\rho_{m0}}{\rho_{X0}}
a^{3(w_{X}-1)} + 1 \right] = \rho_{m0}a^{-3} + \rho_{X0} a^{-3w_{X}}.
\end{equation}

Then, the potential function is given by the integral
\begin{equation}
\label{eq:a4}
V(a)= \int_{0}^{a} \left[ \frac{(3\gamma(a)-2)\kappa_{(4)}^{2}}{6} \rho
+ \frac{(3\gamma(a)-1)\kappa_{(4)}^{2}}{6\lambda} \rho^{2}
+ \frac{2}{3}\sigma^{2} - \frac{\Lambda_{(4)}}{3}
+ \frac{2{\cal U}_0}{\lambda \kappa_{(4)}^{2}} a^{-4} \right] a da,
\end{equation}
where $\rho(a)$ is given by (\ref{eq:a3}) and $\gamma(a)$ comes
from (\ref{eq:a1}).

After substitution the corresponding form of the equation of state
(\ref{eq:a1}), the potential $V(a)$ can be obtained by integration of
(\ref{eq:a4})
\begin{multline}
V(a) = - \frac{\kappa_{(4)}^{2}}{6} \rho_{X0} a^{-3w_{X}+2}
- \frac{\kappa_{(4)}^{2}}{6} \rho_{m0} a^{-1} \\
+ \frac{\kappa_{(4)}^{2}}{6\lambda} \left( \rho_{X0} \rho_{m0} a^{-3w_{X}-1}
- \frac{1}{2} \rho_{X0}^{2} a^{-6w_{X}+2} - \frac{1}{2} \rho_{m0}^{2}
a^{-4} \right) \\ - \frac{\sigma_{0}}{6} a^{-4} - \frac{\Lambda_{(4)}}{6}
a^{2} - \frac{{\cal U}_{0}}{\lambda \kappa_{(4)}^{2}} a^{-2} + \frac{k}{2}.
\end{multline}
It is interesting that in this potential there is a term of the type
$\rho_{X0} \rho_{m0} a^{-3w_{X} - 1}$ which is related to the
co-existence/interaction of both matter and an unknown form of dark energy in the universe.

Obviously the simplest possibility is the pure case of the unknown matter $X$
such that
\[
\rho = \rho_{X0} a^{-3w_{X}}, \quad
p = (w_{X} - 1)\rho, \quad
\gamma(a) = w_{X} = \text{const},
\]
which gives the potential function
\begin{equation}
V(a) = - \frac{\kappa_{(4)}^{2}}{6} \rho_{X0} a^{-3w_{X}+2}
- \frac{\kappa_{(4)}^{2}}{12\lambda} \rho_{X0}^{2} a^{-6w_{X} + 2}
- \frac{\sigma_{0}^2}{6} a^{-4} - \frac{\Lambda_{(4)}}{6} a^{2}
- \frac{{\cal U}_0}{\lambda \kappa_{(4)}^{2}} a^{-2} + \frac{k}{2}
\end{equation}
i.e., $w_{X}$ plays the role of $\gamma$ in (\ref{eq:15}).

As a special example let us consider the case of
$p_{X} = - \frac{2}{3} \rho_{X}$ ($w_{X} = \frac{1}{3}$).
Then, we have
\begin{equation}
V(a) = - \frac{\kappa_{(4)}^{2}}{6} \rho_{X0} a
- \frac{\kappa_{(4)}^{2}}{12\lambda} \rho_{X0}^{2}
- \frac{\sigma_{0}^2}{6} a^{-4} - \frac{\Lambda_{(4)}}{6} a^{2}
- \frac{{\cal U}_{0}}{\lambda \kappa_{(4)}^{2}} a^{-2} + \frac{k}{2}.
\end{equation}

The diagram of the above function is shown in Fig.~\ref{fig:2}.
Note that the second term in $V(a)$ takes the form of an additive
constant like the term $k/2$ and simply scales as curvature term
(in general relativity curvature term appears for $w_X = 2/3$ - cosmic
strings).

\section{Configuration space and stability}

In our further analysis we will explore the dynamics given by the
canonical equations. Assuming $a=x$ and $y=\dot{a}=\dot{x}$, we have from
(\ref{eq:18})
\begin{align}
\label{eq:20}
\dot{x} &= \frac{\partial \mathcal{H}}{\partial y} = y \\
\label{eq:21}
\dot{y} &= - \frac{\partial \mathcal{H}}{\partial x}
= - \frac{\partial V}{\partial x}.
\end{align}
Then, we can perform a qualitative analysis of the system of autonomous differential equations
(\ref{eq:20})--(\ref{eq:21}) in the phase space $(a,\dot{a}) \equiv
(x,y)$.

Firstly, we can observe that trajectories are integrable in quadratures.
Namely, from the Hamiltonian constraint $\mathcal{H} \equiv E = 0$ we obtain
the integral
\begin{equation}
\label{eq:22}
t-t_{0} = \int_{a_{0}}^{a} \frac{da}{\sqrt{-2V(a)}}.
\end{equation}
For some specific forms of the potential function (\ref{eq:15})
we can obviously obtain the exact solutions. However, this is not
the task of our paper since we concentrate on qualitative
analysis only.

\begin{figure}
\includegraphics[width=\textwidth]{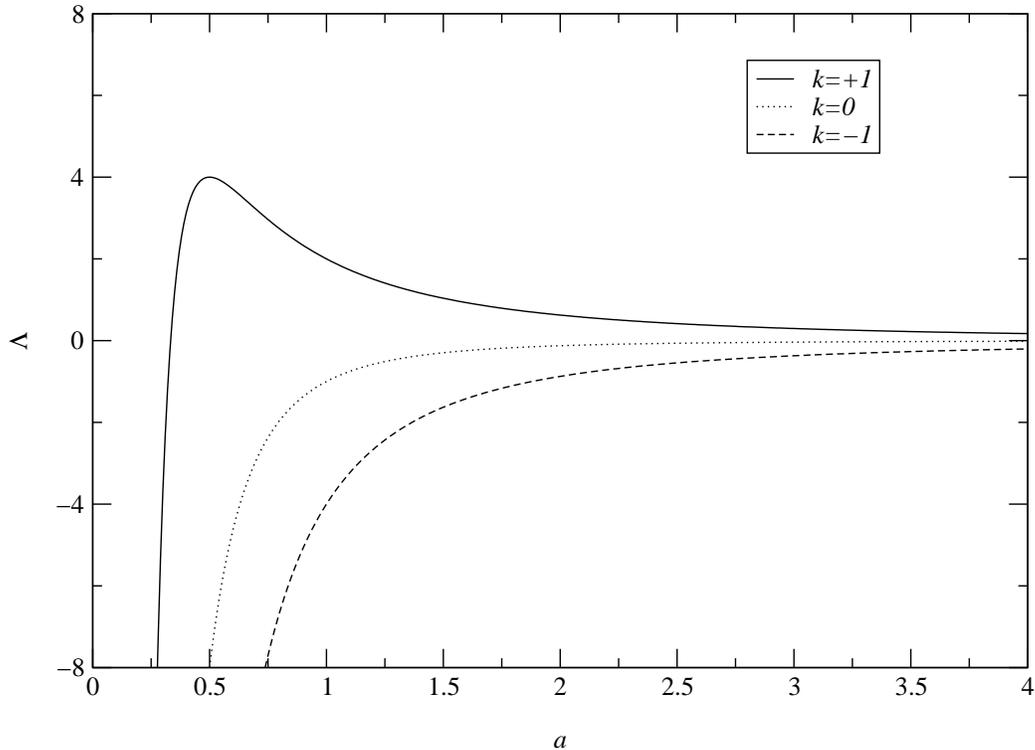}
\caption{The 4-dimensional cosmological constant on the brane,
$\Lambda_{4}(a)$, given by (\ref{eq:25}), for $k=0,-1,+1$ and $\gamma=1$ (dust).
Lines $\Lambda_{(4)}=\text{const.}$ gives qualitative classification of
possible path evolutions. Let us note that there is a similarity
to the standard FRW models. The maximum for $k=+1$ corresponds to a static
universe. The de Sitter model starts at the initial singularity. The Eddington
model starts from the static universe at $t\to -\infty$ (separatrices on the
phase plane) then reaches the singularity (such that $a(t=0)=a_{0}>0$) and
evolves to infinity. The solutions are represented by levels of constant
$\Lambda_{(4)}$ fixed above the $\Lambda_{(4)}(a)$ curve. The domain under
the characteristic curve is non-physical.}
\label{fig:3}
\end{figure}

\begin{figure}
\includegraphics[width=\textwidth]{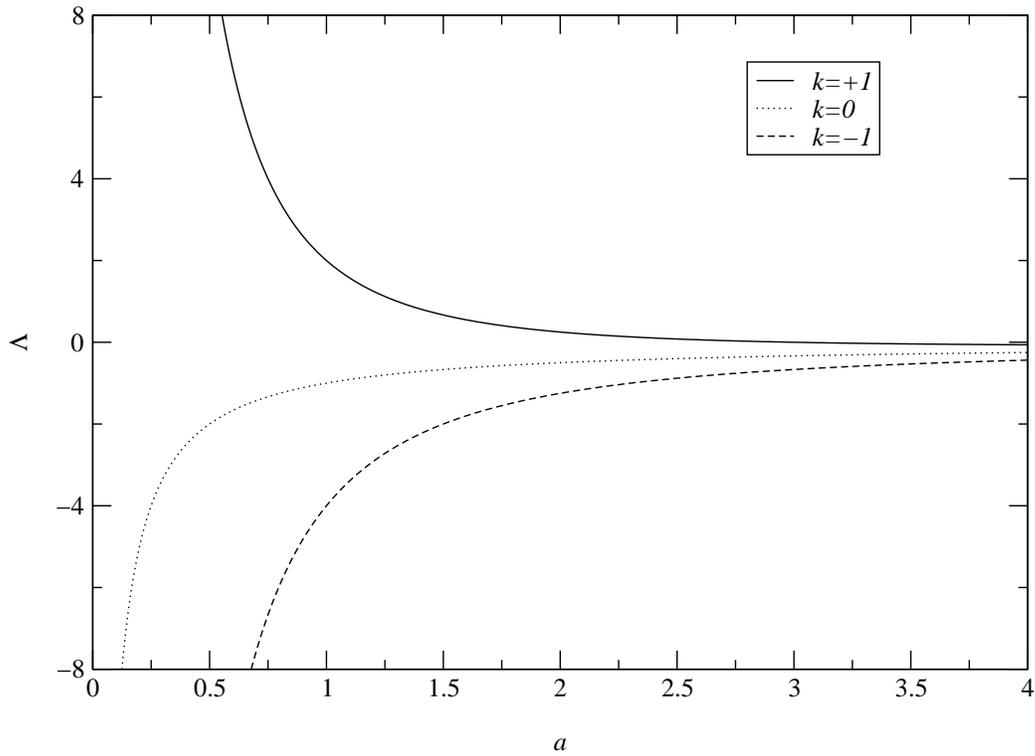}
\caption{The 4-dimensional cosmological constant on the brane,
$\Lambda_{(4)}(a)$, given by (\ref{eq:25}), for $k=0,-1,+1$ and $\gamma=1/3$ (domain walls).
All models with $\Lambda_{(4)}(a)<0$ and $k=0,-1$ oscillate. These with
$k=+1$ are singularity free. The lines of constant $\Lambda_{(4})$ above the
$\Lambda_{(4)}(a)$ curve represents the qualitative evolution of the model.
}
\label{fig:4}
\end{figure}

It is possible to make the classification of qualitative evolution
paths by analyzing the characteristic curve which represents the boundary
equation in the configuration space. For this purpose we consider the
equation of zero velocity, $\dot{a}=0$, which represents the boundary
in the configuration space. Because
\begin{equation}
\label{eq:23}
\dot{a}^{2} = - 2V(a)
\end{equation}
the motion of the system is limited to the region $\{ a\!: V(a)\leq 0\}$.
Consider the boundary of the configuration space given by a condition
\begin{equation}
\label{eq:24}
\partial \mathcal{M} = \{ a\!: V(a) = 0 \}.
\end{equation}
The condition (\ref{eq:24}) together with (\ref{eq:15}) is equivalent to the
equation of the 4-dimensional cosmological constant
on the brane $\Lambda_{(4)}$ expressed as a function of $a$:
\begin{equation}
\label{eq:25}
\Lambda_{(4)}(a) = \frac{1}{a^{2}} \left( - \kappa_{(4)}^{2} \rho_{0}
a^{-3\gamma +2} - \frac{\kappa_{(4)}^{2} \rho_{0}^{2}}{2\lambda}
a^{-6\gamma +2} - \sigma_{0}^2 a^{-4} - \frac{6 {\cal U}_{0}}{\lambda \kappa_{(4)}^{2}}
a^{-2} + 3k \right).
\end{equation}
The plot of $\Lambda_{(4)}(a)$ for different $k$ is shown in
Fig.~\ref{fig:3} ($\gamma=1$) and in Fig.~\ref{fig:4} ($\gamma=1/3$).
We can see that in the latter case ($\gamma = 1/3$ there is no static
Einstein universe.
Finally, we consider the evolution path as a level of $\Lambda_{(4)} =
\text{const} \neq 0$ and then we classify all evolution modulo
their quantitative properties of dynamics (compare
\cite{Dabrowski96,Robertson33,Dabrowski86}). We can conclude that
the classes of admissible models in brane scenario are {\it
qualitatively} the same as in general relativity.

The next advantage of representing dynamics in terms of
Hamiltonian is the possibility to discuss the stability of critical points
which is only based on the geometry of the potential function, namely

--- if a diagram of the potential function $V(a)$ has maxima then
they correspond to {\it unstable} critical points; on the other hand, if
a diagram of the potential function has minima, they correspond to
{\it stable} attractors.

--- if a diagram of the potential function has an inflection point
at $a=a_{0}$, then the corresponding critical point in the phase
plane is a {\it saddle} point.

In general, the stability and the character of a critical point is
determined by the Hessian $[\partial^{2} \mathcal{H}/\partial x^{i}
\partial y^{i}]$.

In our case the Hamiltonian function (\ref{eq:17}) takes the simplest form
for natural mechanical systems (i.e., with the kinetic energy quadratic
in momenta, and the potential energy dependent on generalized coordinates only).
Then, the only possible critical points in a finite domain of phase space
are centers and saddles.

The idea of structural stability originated with Andronov and Pontryagin
\cite{Andronov37}. A dynamical system $S$ is said to be structurally stable
if dynamical systems in the space of all dynamical systems are close,
in the metric sense, to $S$ or are topologically equivalent to $S$.
Instead of finding and analyzing an individual solution of a model,
a space of all possible solutions is investigated. A property is believed to be `realistic' if
it can be attributed to large subsets of models within a space of all
possible solutions or if it possesses a certain stability, i.e., if it
is shared by a slightly perturbed model. There is a wide opinion among
specialists that realistic models should be structurally stable, or even
stronger, that everything should possess a kind of structural stability.
What does the structural stability mean in physics? The problem is
in principle open in more than 2-dimensional case where according to
Smale there are large subsets of structurally unstable systems in the
space of all dynamical systems \cite{Smale80}.
For 2-dimensional dynamical systems, as in the considered case, the
Peixoto's theorem says that structurally stable dynamical systems on
compact manifolds form open and dense subsets in the space of all
dynamical systems on the plane. Therefore, it is reasonable to require
the model of a real 2-dimensional problem to be structurally stable.

When we consider the dynamics of brane world models, then there is a simple
test of structural stability. Namely, if the right-hand sides of the
dynamical systems are in polynomial form, the global phase portraits
are structurally stable $S^2$ ($\mathbb{R}^2$ with a Poincar{\'e} sphere)
if and only if the number of critical points and limit cycles is
finite, each point is hyperbolic and there are no trajectories connecting saddle
points. In the considered case the points at infinity are revealed on
the projective plane. Two projective maps $(z,u)$, $(v,w)$ cover a circle
at infinity given by $z=0$ ($z=\infty$) and $v=0$ ($y=\infty$).

Therefore, one can conclude that the {\it brane-world models are structurally
stable}. It holds because the potential function $V(a)$ is convex up and then
there are no non-hyperbolic centres. There are also no trajectories
connecting saddle points. All these properties can be deduced from
the geometry of the potential function $V(a)$.

Structural stability is sometimes considered as a precondition of the
`real existence'. Having too many drastically different mathematical
models - all of them equally well fitting the observational data
(up to measurement errors) - seems to be fatal for
the empirical method of modern science \cite{Thom77}. Therefore, any
2-dimensional structurally unstable model is not of physical
importance.

From the physical point of view it is interesting to answer the
question: are the trajectories distributed in the phase space in such
a way that critical points are typical or exceptional? How are
trajectories with interesting properties distributed? For example,
along which trajectories the acceleration condition, $\ddot{a} =
- dV/da > 0$, is satisfied? One can easily decuce this from the geometry
of the potential function. In the phase space, the {\it area of acceleration}
is determined by $\dot{y}>0$ or by the condition that
\begin{equation}
\label{eq:26}
\frac{(3\gamma -2)\kappa_{4)}^{2} \rho_{0}}{6} a^{-3\gamma} +
\frac{(3\gamma -1)\kappa_{(4)}^{2}\rho_{0}^{2}} a^{-6\gamma} +
\frac{2}{3} \sigma_{0} a^{-6} -\frac{\Lambda_{(4)}}{3}a +
\frac{2{\cal U}_{0}}{\lambda \kappa_{(4)}^{2}} a^{-4} < 0.
\end{equation}
From (\ref{eq:26}) we can see that in order to obtain acceleration, the
{\it positive} 4-dimensional cosmological constant $\Lambda_{(4)}$ is necessary.

It can easily be demonstrated that if $\dot{a}(t) \to \text{const}$
as $a \to 0$, then the corresponding world model has {\it no particle
horizon}. Indeed, if there exists a constant $C$ such that for
sufficiently large $\epsilon$, $da/dt < C$, then
\begin{equation}
\label{eq:27}
\int_{0}^{a_{0}} \frac{da}{a} < C \int_{0}^{t_{0}} \frac{dt}{a}
= C(\eta_{0} - \eta_{\text{sing}}).
\end{equation}
The integral on the left hand-side of this formula diverges which
means that the time $\eta$ goes to minus infinity, and that there
are no causally disconnected regions. Putting this in terms of
variables $\{x,y\}$ one needs $x \to 0$, $y \to \text{const}$ for
the horizon problem to be solved. Now, from the Hamiltonian
constraint we obtain that as $x\to 0$, then $V(x) \to \text{const}$.
Therefore, if the {\it horizon problem is solved}, then the curvature
effects cannot be neglected in the vicinity of an initial singularity.

\section{Phase plane analysis of the models with $\gamma = 1$ and $\gamma = 1/3$}

In this section we discuss the dynamics of the brane universes in a more detailed way.
Firstly, it is easy to verify that the general relativistic limit can be recovered if
one takes
\[
\frac{1}{\lambda} \to 0
\]
in Eq. (\ref{eq:11}). The quadratic contribution of the brane $\rho^{2}$
matters only when \cite{roy01a}
\[
\rho > \lambda > (100 \text{ GeV})^{4}.
\]
In order to simplify Eq. (\ref{eq:11}) one can assume that $\kappa_{(4)}^{2}=1$ or
${}^{(4)}M_{p}^{2} = 1/8\pi = (1.2 \times 10^{19} \text{GeV})^{2}$
which gives
\[
\lambda = \frac{6}{\kappa_{5}^{4}}, \qquad
\Lambda_{(4)} = \frac{1}{2} \kappa_{5}^{2} \Lambda_{(5)} + \frac{1}{2}
\lambda = \frac{1}{2} \sqrt{\frac{6}{\lambda}}\Lambda_{(5)} +
\frac{1}{2} \lambda
\]
and finally Eq. (\ref{eq:11}) reads as
\begin{equation}
\label{eq:28}
H^{2} = \frac{\rho}{3} \left( 1 + \frac{\rho}{2\lambda} \right)
+ \frac{1}{6} \left( \sqrt{\frac{6}{\lambda}}\Lambda_{(5)} +
\frac{1}{2} \lambda \right) + \frac{\sigma_0^2}{6} a^{\-6}
- k a^{-2} + \frac{2{\cal U}_{0}}{\lambda} a^{-4}.
\end{equation}
Therefore, the dynamics depends on four independent constants. Without
losing a generality, we put $\Lambda_{(5)} = {\cal U}_{0} = \rho_{0}=1$. Finally,
the dynamics depends on one parameter $\lambda$ (brane tension) only.

As a proof of the effectiveness of the presented method we consider only
two limit subcases: pure dust matter and pure matter $X$.

\begin{figure}
\includegraphics[width=\textwidth]{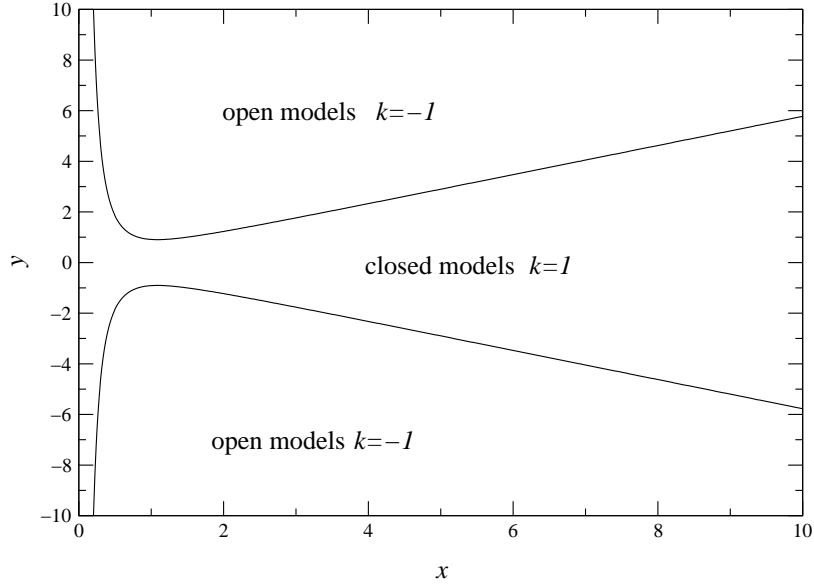}
\caption{The division of phase space for system (\ref{eq:29})-(\ref{eq:30})
on different domains with respect to the curvature index. The flat model
trajectory $k=0$ separates the regions of the models with negative and
positive curvature.}
\label{fig:5}
\end{figure}

Then, the dynamical system (\ref{eq:20})-(\ref{eq:21}) has the form
\begin{align}
\label{eq:29}
\dot{x} &= y \\
\label{eq:30}
\dot{y} &= - \frac{3\gamma -2}{6}x^{-3\gamma +1}
- \frac{3\gamma -1}{6\lambda} x^{-6\gamma +1} - \frac{2}{3} \sigma_{0}x^{-5}
+ \frac{\Lambda_{(4)}}{3} x - \frac{2}{\lambda}x^{-3}
\end{align}
where
$\Lambda_{(4)} = \frac{1}{2} \sqrt{\frac{6}{\lambda}} + \frac{1}{2} \lambda$.
We can see that the vector field in (\ref{eq:30}) is not smooth.
There are two important cases: $\sigma_{0}=1$ and $\sigma_{0}=0$.
The former describe the Bianchi models (Bianchi I for $k=0$ and
Bianchi V for $k=-1$). The latter describes the FRW models ($k=0,\pm 1$).
The qualitative dynamics of both is similar.
Due to the existence of the energy integral (\ref{eq:18}), the phase space
is separated into the two domains by the flat model trajectory (Fig.~\ref{fig:5}).
\begin{figure}
\includegraphics[angle=-90,width=\textwidth]{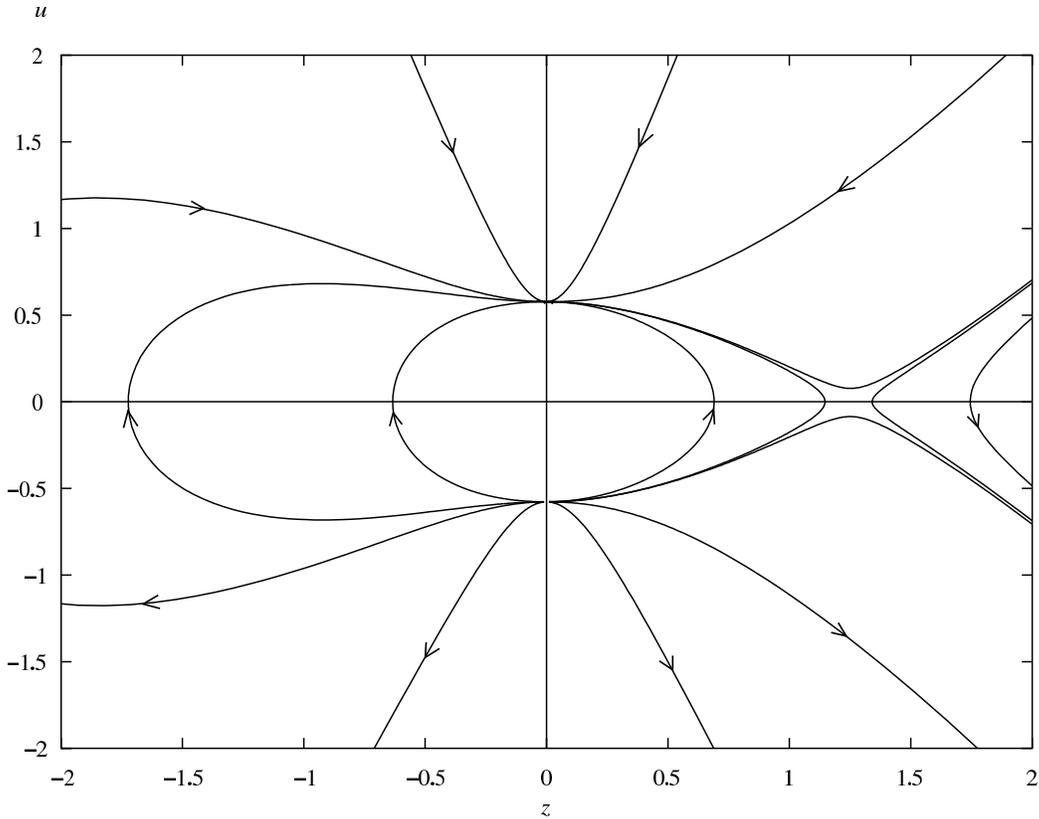}
\caption{The phase portrait of the system (\ref{eq:29})--(\ref{eq:30}) with
$\lambda=10^{8}$ and $\gamma=1$. The Einstein-de Sitter and Eddington models
are represented by separatrices in the phase plane. In the neighbourhood
of separatrices one can see the Lema{\^\i}tre-Eddington (L-E) evolution with
characteristic quasi-static regime. The closer to a critical point
the trajectory is, the longer the time of a quasi-static stage evolution.
All phase curves lie on algebraic curves given by the first intergral
(Hamiltonian constraint). The acceleration region is situated to the right of
the saddle point, therefore for the L-E universes the acceleration begins in
the middle of quasi-static stage.}
\label{fig:pf1}
\end{figure}
\begin{figure}
\includegraphics[angle=-90,width=\textwidth]{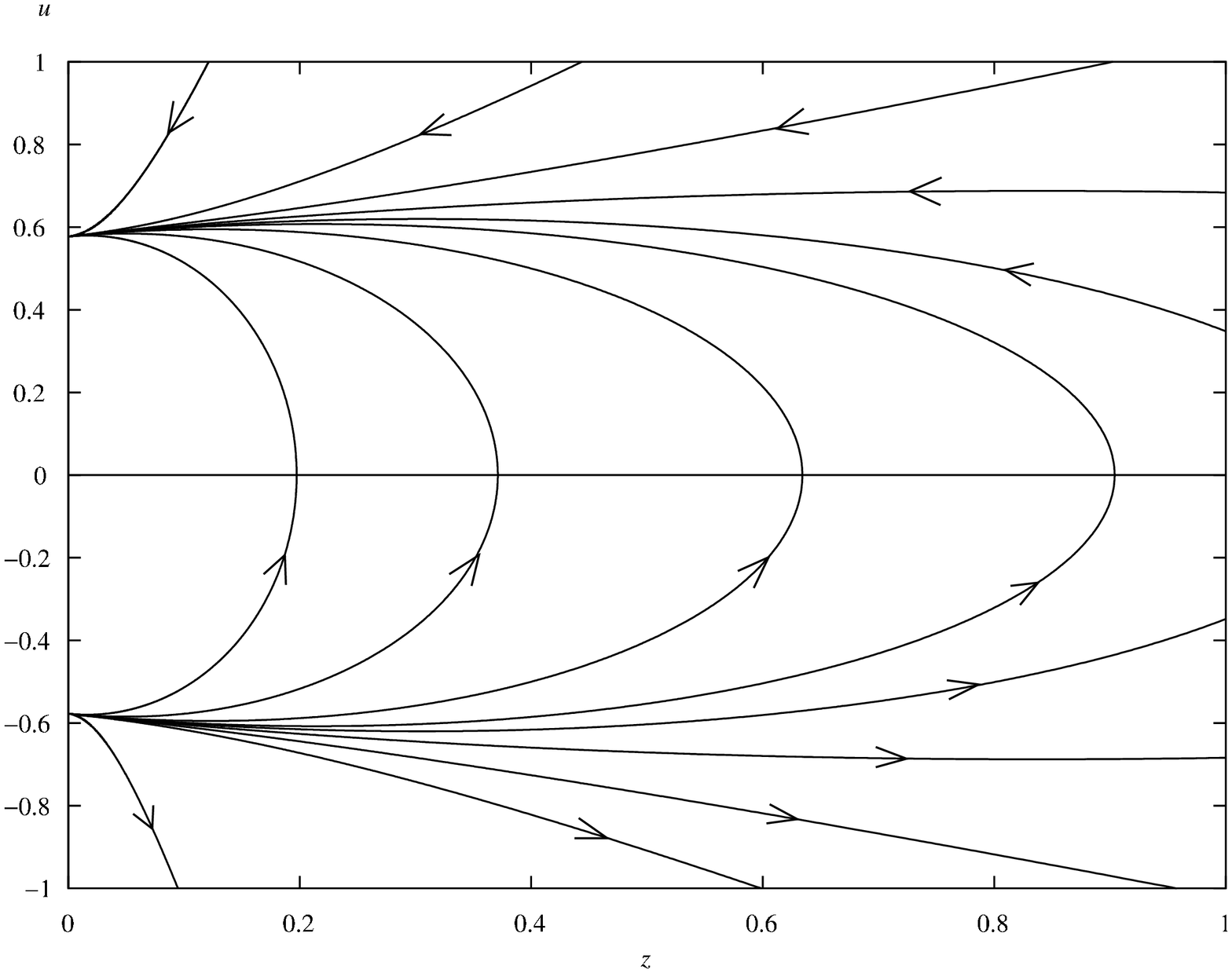}
\caption{The phase portrait of system (\ref{eq:29})--(\ref{eq:30}) with
$\lambda=10^{8}$ and $\gamma=1/3$. The negative curvature term and the
brane term in potential (\ref{eq:15}) are of the same type. There are
two types of trajectories which expand to the maximal scale factor $a$ and
then recollapse or vice versa. In the first type of evolution the closed
models start with the singularity $a(0)=0$ and $\dot{a}=\infty$, reach
the maximum radius and recollapse while the other class of models
recollapse from $a(0)=\infty$ and $\dot{a}=-\infty$ to the minimum
value of $a$, and then expand to infinity. As in the $\gamma =1$ case,
the acceleration region is on the right to the saddle point.}
\label{fig:pf2}
\end{figure}
Let us consider first the case of the FRW dynamics with dust $\gamma =1$.
In this case the effects of brane and shear in the Bianchi I models are
equivalent. An interesting result which shows that there exists a maximum
of scalar shear in the Bianchi I model was obtained by Toporensky
\cite{Toporensky01}.

\begin{figure}
\includegraphics[angle=-90,width=\textwidth]{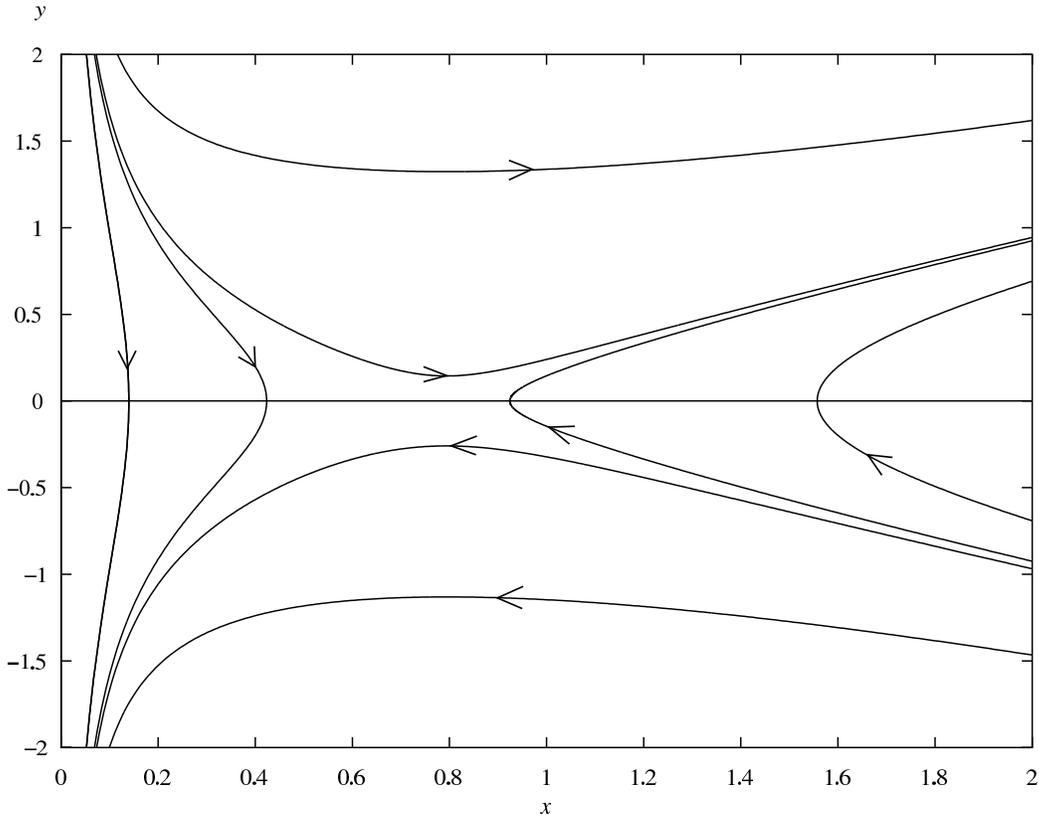}
\caption{The phase portrait of system (\ref{eq:31})--(\ref{eq:32}) with
$\lambda=10^{8}$ and $\gamma=1$. All closed generic models are of two types.
The first starts from the anti-de Sitter and finishes at de Sitter stage.
The second starts from an initial singularity, expands to a maximum size,
and then recollapses to a second singularity. Non-generic (exceptional)
cases corresponds to the separatrices going in or out of the saddle point,
which represent the static universe. For all open and expanding universes
the de Sitter model is a global attractor, and for open and contracting
universes the anti-de Sitter is a global repeller. The singularity is
reached at $z=\infty$ and the half-plane $z<0$ has no physical sense.
All points at the infinity which are represented by $\{ z=0 \}$ axis
are hyperbolic, therefore the system is {\it structurally stable}. The same
result is valid in the $(v,w)$ plane. The brane effects produce
$\rho^2/\lambda$ term which is formally equivalent to the effects
of the stiff matter equation $p=\rho$ (or a massless scalar field).}
\label{fig:pf3}
\end{figure}

The system (\ref{eq:29})-(\ref{eq:30}) can be regularized at the origin, after
introducing the projective coordinates
\begin{equation}
z=\frac{1}{x}, \qquad u=\frac{y}{x}.
\end{equation}
Then, we obtain
\begin{align}
\label{eq:31}
\dot{z} &= - u z \\
\label{eq:32}
\dot{u} &= -\frac{1}{6}z^{3} - \frac{2}{3\lambda} z^{6} +
\frac{\Lambda_{(4)}}{3} - \frac{2}{\lambda} z^{4} - u^{2}.
\end{align}
The system has only three critical points in this map
\begin{align*}
u_{0} &= \pm \sqrt{\frac{\Lambda_{4}}{3}},& z_{0}&=0, \\
u_{0} &=0,& z&=z_{0} \!: 2z^6 + 12z^4 + \lambda z^3 - 2\Lambda_{(4)} \lambda=0.
\end{align*}
In order to find that there is one critical point with $z_{0} > 0$, it is
sufficient to consider the diagram of the function
\[
f(z) = 2z^6 + 12z^4 + \lambda z^3 - 2\Lambda_{(4)} \lambda.
\]
If $\Lambda_{(4)} \lambda >0$ then $f(0) < 0$ and because $f(z)$ is
strictly increasing as $z>0$ then there exits
only one critical point such that $z_{0}>0$, such that $f(z_{0})=0$.
Let us note that
for $\Lambda_{(4)}<0$ there is no such a point. The phase portrait of this
system is presented in Fig.~\ref{fig:pf3}.

\begin{figure}
\includegraphics[angle=-90,width=\textwidth]{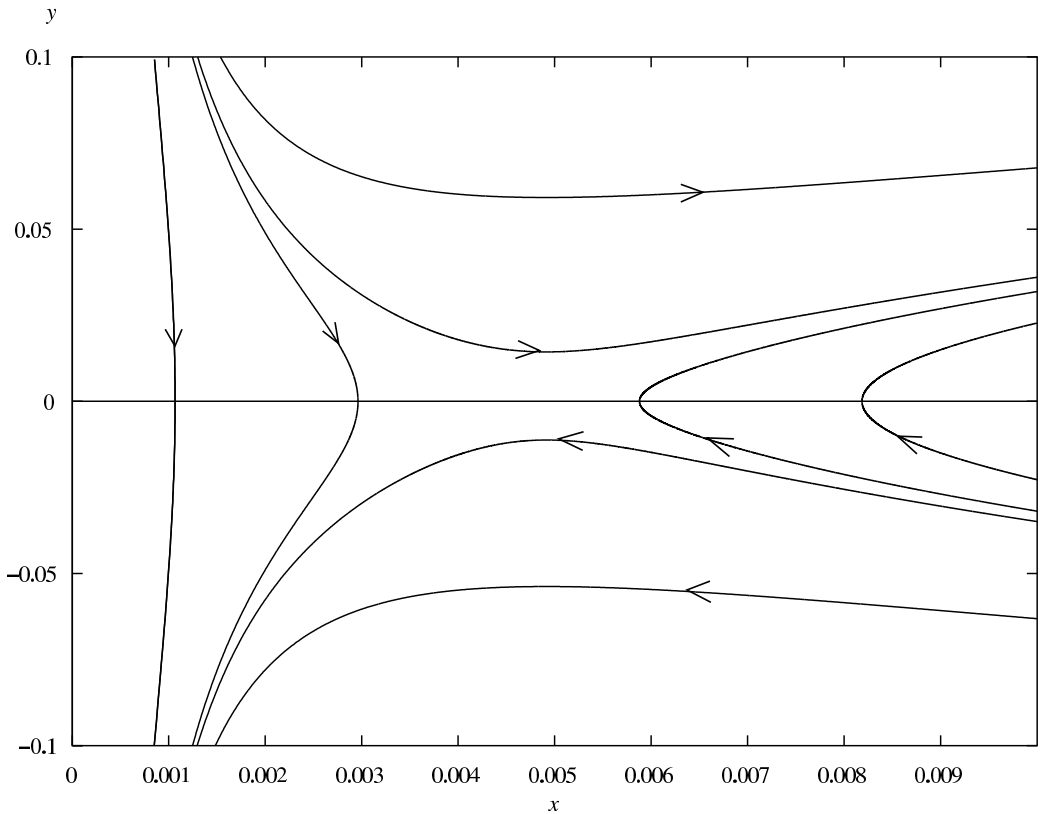}
\caption{The phase portrait of system (\ref{eq:37})--(\ref{eq:38}) with
$\lambda=10^{8}$ and $\gamma=1/3$. The acceleration region is situated
on the left from the saddle point. Therefore the recollapsing and then
expanding models lie permanently in the acceleration domain. The
Eddington model is also in this region. The Lema{\^\i}tre-Eddington
type models start accelerating in the middle of quasi-static phase.}
\label{fig:pf4}
\end{figure}

For the completeness of the analysis of the dynamical system,
it is necessary to consider another projective map $(v,w)$
\begin{equation}
\label{eq:33pre}
v=\frac{1}{y}, \qquad w=\frac{x}{y}.
\end{equation}
After regularization and after introducing a new time variable
$\eta\!: d\eta = dt/w^{5}$ we obtain from (\ref{eq:29})-(\ref{eq:30}) and
(\ref{eq:33pre})
\begin{align}
\label{eq:33}
\dot{v} &= v \left( \frac{1}{6} v^{3} w^{3} + \frac{1}{3\lambda} v^{6}
- \frac{\Lambda_{(4)}}{3} w^6 + \frac{2}{\lambda} v^{4} w^{2} \right) \\
\label{eq:34}
\dot{w} &= w \left( w^{4} + \frac{1}{6} v^3 w^3 + \frac{1}{3\lambda} v^{6}
- \frac{\Lambda_{(4)} }{3} w^{6} + \frac{2}{\lambda} v^{4} w^{2} \right).
\end{align}
In this map we can find only one critical point $v=0$ and $w=0$.
As it was mentioned before, it is interesting to observe how much
time the trajectories spend in the area of accelerated expansion defined as
\begin{gather*}
\{ (x,y)\!: x > 0 \wedge \dot{y} > 0 \} \Leftrightarrow \\
\left\{ (x,y)\!: \frac{3\gamma -2}{6}x^{-3\gamma} + \frac{3\gamma -1}{6\lambda}
x^{-6\gamma} + \frac{2}{3} \sigma_{0} x^{-6} - \frac{\Lambda_{(4)}}{3}
+ \frac{2}{\lambda} x^{-4} < 0 \right\}.
\end{gather*}
For the special case of dust we obtain in $(z,u)$-coordinates
\[
D_{\text{accel}} = \left\{ (z,u)\!: \frac{1}{6}z^{3} + \frac{1}{3\lambda}
z^{6} + \frac{2}{3}\sigma_{0} z^{6} - \frac{\Lambda_{(4)}}{3}
+ \frac{2}{\lambda} z^{4} > 0 \wedge z>0 \right\}.
\]

The dynamical system for pure accelerating matter has the form
analogous to (\ref{eq:29})--(\ref{eq:30}) but now $w_{X}$ coincides
to $\gamma$. As an illustration we consider the case of
$p_{X} =-\frac{2}{3} \rho_{X}$ (domain-wall-like matter), i.e.
$w_{X}=\frac{1}{3}$. Then, we have
\begin{align}
\label{eq:35}
\dot{x} &= y \\
\label{eq:36}
\dot{y} &= \frac{1}{6} - \frac{2}{\lambda} x^{-3} +
\frac{\Lambda_{(4)}}{3}x.
\end{align}
In this case, the term which comes from the brane and which is proportional
to $\rho^{2}$ vanishes (see Eq.~\ref{eq:11}), and only the term which comes
from a non-vanishing Weyl tensor is present in (\ref{eq:35})-(\ref{eq:36}).

After introducing the projective coordinates
\[
z=\frac{1}{x}, \qquad u=\frac{y}{x}
\]
we obtain
\begin{align}
\label{eq:37}
\dot{z} &= - uz \\
\label{eq:38}
\dot{u} &= \frac{1}{6}z - \frac{2}{\lambda} z^{4}
+ \frac{\Lambda_{(4)}}{3} - u^{2}.
\end{align}
The critical points are
\begin{align*}
u_{0}&=\pm\sqrt{\frac{\Lambda_{(4)}}{3}},& z_{0}&=0, \\
u_{0}&=0,& z_{0}&=z\! : f(z)=\frac{1}{6}z - \frac{2}{\lambda}z^{4}
+\frac{\Lambda_{(4)}}{3}=0.
\end{align*}
It can be shown that $f(z_{0})=0$ always exists because $f(0)>0$ and
$f(\infty) = -\infty$.

In turn, in the map
\[
v=\frac{1}{y}, \qquad w=\frac{x}{y}
\]
we have
\begin{align}
\frac{dv}{d\eta} &= w^{3} \frac{dv}{dt} = -v \left( \frac{1}{6} v w^{3}
+ \frac{\Lambda_{(4)}}{3} w^{4} - \frac{2}{\lambda} v^{4} \right) \\
\frac{dw}{d\eta} &= w^{3} \frac{dw}{dt} = w^{3} - w \left( \frac{1}{6}
v w^{3} + \frac{\Lambda_{(4)}}{3} w^{4} - \frac{2}{\lambda} v^{4}
\right).
\end{align}

From (\ref{eq:35})--(\ref{eq:36}) one can easily determine the domain
of acceleration which corresponds to the domain of $x$ in which
$V(x)$ is a decreasing function of its argument
\[
- \frac{dV}{dx} > 0 \quad \Leftrightarrow \quad x > x_{\text{max}}
\]
where $x_{\text{max}}\!: \frac{\partial V}{\partial x}|_{x=x_{\text{max}}} = 0$ or
$\dot{y}(x_{\text{max}}) = 0$. It can be checked that $\dot{y}$ is a strictly
increasing function of $a$ ($\ddot{y}>0$) and $\dot{y}(0)=-\infty$,
so there is a single point $x_{\text{max}}$.

From (\ref{eq:36})
we can also observe that as $x$ goes to zero then $y$ goes to
infinity, which means that the model has a particle horizon
($\Lambda_{(4)}$ is negligible and then $y \propto x^{-1}$). It is due to
the presence of $u_{0}\neq0$.

From (\ref{eq:29})--(\ref{eq:30}) and the integral of energy we can see
that we have the horizon in a generic case if $\rho_{0}\neq 0 \wedge
u_{0}\neq 0 \wedge \sigma_{0}\neq 0$. In the special case of
$u_{0}=\sigma_{0}=0$ we have a model without a horizon (the FRW models
with vanishing electromagnetic part of Weyl tensor).

From (\ref{eq:35})--(\ref{eq:36}), we obtain that as $x \to 0$, then
$\dot{y}=\frac{1}{6}$, i.e. $y \propto t$ and $x(t) \propto t^{2}$.

In general, from the first integral (\ref{eq:18}) one observes that
the generic evolution of cosmological models without a horizon is
when $V(a) \to \text{const.}$ (zero is also possible) as $a \to 0$.

Effects of $\Lambda_{(4)}$ are always negligible near a singularity
and if $u_{0}\neq 0$ we have always asymptotically $V(0)=\infty$.
For $u_{0}=\sigma_{0}=0$ and $\gamma \le \frac{1}{3}$ (domain walls on the
brane), then $V(a)$ goes
to a constant and there is no horizon in the model. Then,
the generic behaviour near the singularity is $x(t) \propto t$,
i.e., it is Milne's evolution (in general relativity this appears for
$\gamma = 2/3$ (cosmic strings on the brane \cite{Dabrowski96}).

\section{Simple dynamics in terms of density parameters $\Omega$ and
the exact solutions}

Independent observations of supernovae type Ia, made by the Supernovae
Cosmology Project and the High $z$ Survey Team \cite{Supernovae},
indicate that our Universe is currently accelerating. There is a fundamental
problem for theoretical physics
to explain the origin of this acceleration. If we introduce the
cosmological constant $\Lambda_{(4)}$ and assume that the Universe is
flat, then the best-fit model is for the cosmological constant density parameter
$\Omega_{\Lambda_{(4)},0}=0.72$ and for the dust density parameter
equal to $\Omega_{m,0}=0.28$ (index "0" refers to the present moment of time).

Our formalism gives natural base to express dynamical equations in
terms of dimensionless observational density parameters $\Omega$ and to compare
the results with supernovae data.
However, before we study these quantities in detail following the discussion
of Refs. \cite{Dabrowski96,Dabrowski86,AJI,AJIII} we introduce
the notation in which it is easy to tell which models are exactly
integrable. The Friedmann Eq. (\ref{eq:11}) with the help of the conservation equations
(\ref{eq:12}) and (\ref{eq:13}), can be rewritten in the form
\begin{equation}
\label{FriedCCC}
\frac{1}{a^2} \left( \frac{da}{dt} \right)^2 =
\frac{C_{GR}}{a^{3\gamma}} + \frac{C_{\lambda}}{a^{6\gamma}} -
\frac{k}{a^2} + \frac{\Lambda_4}{3} + \frac{C_{\cal U}}{a^4} ,
\end{equation}
where we have defined the appropriate constants ($\kappa_{(4)}^2 = 8\pi G$)
\bea
C_{GR} & = & \frac{\kappa_{(4)}^2}{3}a^{3\gamma} \varrho ,\\
C_{\lambda} & = & \frac{\kappa_{(4)}^2}{6\lambda} a^{6\gamma}
\varrho^2 ,\\
C_{\cal U} & = & \frac{2}{\kappa_{(4)}^2 \lambda}
a^4 {\cal U} ,
\eea
and $C_{GR}$ is a of general relativistic nature. It is easy to notice that the
following cases can be exactly integrable in terms of elliptic
functions \cite{Dabrowski86,Dabrowski96}: $\gamma = 0$
(cosmological constant), $\gamma = 1/3$ (domain-wall-like matter)
and $\gamma = 2/3$ (cosmic strings). The first case is the easiest
since in this case the first two terms on the right-hand-side of
(\ref{FriedCCC}) play the role of cosmological constants similar
to $\Lambda_{(4)}$. The next two cases involve terms which were already integrated
in the context of general relativity. For $\gamma = 1/3$ (domain-wall-like matter on the
brane) the general relativistic term with $C_{GR}$ in (\ref{FriedCCC}) scales as domain-walls
in general relativity while the term with $C_{\lambda}$ scales as cosmic string (curvature)
in general relativity. For $\gamma = 2/3$ the term with $C_{\lambda}$ scales as radiation
in general relativity. Then, the problem of writing exact solutions
reduces to the repetition of the discussion of Refs.
\cite{Dabrowski86,Dabrowski96}. We will not be doing this here.
For other values of $\gamma = 4/3; 1; 2$ the terms of the type $1/a^8$
and $1/a^{12}$ appear and the integration involves hyperelliptic
integrals.

Coming back to observational quantities we now define the
dimensionless observational density parameters \cite{AJI,AJII,AJIII}
\bea
\label{Omegadef}
\Omega_{GR} & = & \frac{\kappa_{(4)}^2}{3H^2} \varrho ,\\
\Omega_{\lambda} & = & \frac{\kappa_{(4)}^2}{6H^2\lambda} \varrho^2 ,\\
\Omega_{\cal U} & = & \frac{2}{\kappa_{(4)}^2 H^2\lambda} {\cal U}
,\\
\Omega_{k} & = & - \frac{k}{H^2a^2} ,\\
\Omega_{\Lambda_{(4)}} & = & \frac{\Lambda_{(4)}}{3H^2} ,
\eea
where the Hubble parameter and the deceleration parameter read as
\bea
H & = & \frac{\dot{a}}{a} ,\\
q & = & - \frac{\ddot{a}a}{\dot{a}^2} ,
\label{HaQu}
\eea
so that the Friedmann equation (\ref{eq:11}) can be written down
in the form
\begin{equation}
\label{Om=1}
\Omega_{GR} + \Omega_{\lambda} + \Omega_{k} + \Omega_{\Lambda_{(4)}} + \Omega_{\cal U}
= 1  .
\end{equation}
Using (\ref{Omegadef})-(\ref{HaQu}) the equation (\ref{eq:13}) can now be
rewritten as (compare Eq.(10) of \cite{AJI})
\begin{equation}
\label{Lambda4}
\frac{\Lambda_{(4)}}{3H^2} = \frac{3\gamma - 2}{2} \Omega_{GR} +
\frac{3\gamma - 1}{2} \Omega_{\lambda} + \Omega_{\cal U} - q .
\end{equation}
On the other hand, it is useful to express the curvature of spatial sections
by observational quantities using (\ref{Om=1}) and (\ref{Lambda4})
\begin{equation}
\label{indexk}
\frac{k}{H^2a^2} = \frac{3\gamma}{2} \Omega_{GR} +
\frac{3\gamma + 1}{2} \Omega_{\lambda} + 2\Omega_{\cal U} - q - 1.
\end{equation}
These relations (\ref{Lambda4}) and (\ref{indexk}) together with
the exact solutions (which here are certainly available for $\gamma = 0; 1/3;
2/3$) (see Refs. \cite{AJIII,IAU87}) are useful in writing down an explicit redshift-magnitude
formula (generalized Hubble law) for the brane models to study
their compatibility with astronomical data from supernovae. Such a
comparison will be presented elsewhere \cite{SzDabKr02}.

In this paper we only study the formalism which allows to
formulate our phase space quantities in terms of the observational
parameters $\Omega$. Let us consider a brane universe filled with an unknown
type of matter with barotropic equation of state
$p_{i}= (w_{i} - 1)\rho_{i}$. Then, it is useful to rewrite the dynamical equations to
a new form using dimensionless quantities
\begin{equation}
\label{eq:b1}
x \equiv \frac{a}{a_{0}}, \qquad
T \equiv |H_{0}| t .
\end{equation}

The basic dynamical equations are then rewritten as
\begin{align}
\label{eq:b4}
\frac{\dot{x}^{2}}{2} &= \frac{1}{2} \Omega_{k,0} + \frac{1}{2}
\sum_{i} \Omega_{i,0} x^{2-3w_{i}} = - V(x) \\
\label{eq:b5}
\ddot{x} &= \frac{1}{2} \sum_{i} \Omega_{i,0} (2-3w_{i})x^{1-3w_{i}} ,
\end{align}
where $\Omega_i = (\Omega_{GR}, \Omega_{\lambda}, \Omega_{\cal U},
\Omega_{\Lambda_{4}})$.

The above equations can be represented as a two-dimensional dynamical system
\begin{align*}
\dot{x} &= y \\
\dot{y} &= \frac{1}{2} \sum_{i} \Omega_{i,0} (2-3w_{i}) x^{1-3w_{i}}.
\end{align*}
As we have mentioned already the dynamics can be always reduced to the form of the
Friedmann models of general relativity with a certain kind of a non-interacting multifluid.

Of course, the dynamical system (\ref{eq:b4})-(\ref{eq:b5}) has the Hamiltonian
\begin{equation}
\label{eq:b6}
\mathcal{H} = \frac{p_{x}^{2}}{2} + V(x),
\end{equation}
where $V(x) = - \frac{1}{2} \Omega_{k,0} - \frac{1}{2} \sum_{i}
\Omega_{i,0} x^{2-3w_{i}}$, which should be considered on the zero-energy
level.

As an example of application of these equations to study the problem of cosmic
acceleration, consider the case of
$\Omega_{{\cal U},0}=0$ and $\Omega_{\Lambda_{(4)},0} \ne 0$. It emerges that at present
our Universe accelerates provided that
\begin{equation}
\label{eq:b7}
\Omega_{X,0}(2-3w_{X})x^{1-3w_{X}} +
\Omega_{\lambda,0}(2-6w_{X})x^{1-6w_{X}} -
2\Omega_{\Lambda_{(4)},0}x > 0 ,
\end{equation}
where $w_X$ refers to any type of matter $X$ with the equation of
state $p_X = (w_X - 1) \varrho_X$ (for $\Omega_{\Lambda_{(4)}} = 0$
it is possible if $w_{X} < 1/3$).

It is convenient to introduce a new variable $z=x^{-3w_{X}} > 0$,
then Eq. (\ref{eq:b7}) reduces to the quadratic inequality
\begin{equation}
\label{eq:b8}
\Omega_{\lambda,0}(2-6w_{X})z^{2}
+ \Omega_{X,0}(2-3w_{X})z
- 2\Omega_{\Lambda_{(4)},0} > 0.
\end{equation}

In the special case of the component $X$ in the form of the wall-like-matter
$w_{X} = 1/3$, the Universe accelerates if
\begin{equation}
\label{eq:b9}
\Omega_{X,0}z > - 2\Omega_{\Lambda_{(4)},0} ,
\end{equation}
i.e., if $\Lambda_{(4)} > 0$, $w_{X}>0$, the universe always accelerates.

In general, the domain of acceleration depends on the solution of
the inequality (\ref{eq:b8}). We assume that we are in the region of
acceleration $\Omega_{\Lambda_{(4)},0} > 0$, $w_{X}>0$ and consider two cases.

\noindent Case 1: $w_{X}>1/3$. \\
There are always two solutions for $z$ with opposite signs.
One solution is for $0<z<z_{+}$ (or
$x_{+}<x<\infty$), where $x_{+}$ is given by
\begin{equation}
x_{+} = \left[\frac{(3w_{X}-2)\Omega_{X,0} +
\sqrt{[(2-3w_{X})\Omega_{X,0 }]^2 -16(1-3w_{X})\Omega_{\Lambda_{(4)},0}
\Omega_{\lambda,0}}}{(3w_{X}-1)
\Omega_{\lambda,0}}\right]^{-1/3w_{X}}.
\end{equation}
In this case we have a minimum value of acceleration at
\begin{equation}
x_{\mathrm{min}} = \left[\frac{(2-3w_{X})\Omega_{X,0}}{2(6w_{X}-2)
\Omega_{\lambda,0}}\right]^{-1/3w_{X}}.
\end{equation}

\noindent Case 2: $w_{X} < 1/3$. \\
The universe accelerates for every $x$.

Let us note that in Case 1 the minimum value of $x$ can always be
expressed in terms of redshift
$z=x^{-1}-1$ and from the inequality $x_{+} \le 1$ we obtain
an additional restriction on parameters $\Omega_{i,0}$.

\section{Conclusion}

In the paper we studied the dynamics of Randall-Sundrum brane
universes with isotropic Friedmann geometry. Our approach is the
simplest in that we formulate the problem in a 2-dimensional phase
space and not in a higher-dimensional phase space as it has been
done recently. First of all, this is the smallest possible
dimension to study the isotropic cosmological systems of
equations. Then, it allows to {\it avoid} the degeneracy of
critical points and trajectories which appear in a higher
dimensional phase space and so to {\it avoid} structural
instability of these models. Also, since the system is Hamiltonian
one is able to study its properties in a one-dimensional
configuration space rather than in phase space. In such a space
the motion of the universe point is reduced to the motion of a
particle with the potential energy $V(a)$, where $a$ is the scale
factor (see Section 2). On the other hand, Hamiltonian formulation
can easily be applied to quantum cosmology on the brane.

In Fig.~\ref{fig:1} and \ref{fig:2} we can see that there is no
qualitative difference in the shapes of the potential function $V(a)$
as $\gamma$ varies for cases under studies. Only the values of
$a$ for which the potential is negative have the physical meaning.
The maximum of the potential function corresponds to an unstable
saddle point on a phase diagram (a critical point solution of the
zero energy level). The de Sitter solution is a point at
infinity. Near the initial singularity we have a solution
$a(t) \propto t^{1/3\gamma}$ which differs from a general relativistic one
$a(t) \propto t^{2/3\gamma}$ .

From the theory of qualitative differential equations in Sections 3 and 4 we
obtained the visualization of the system evolution in the phase plane $(x,\dot{x})$
and analyzed the asymptotic states and concluded that the brane
models are {\it structurally stable}. We studied the solution of the
(past) horizon problem and the initial conditions for acceleration of
our Universe. We have
the neat interpretation of {\it a domain of acceleration} as a domain in
configuration space where the potential function decreases. Therefore
from the observation of the potential function $V(a)$, we can see the
acceleration domain $a>a_{\text{max}}$, with $V'(a_{\text{max}})=0$,
which is independent of the curvature index $k$. On the other
hand, the (past) horizon problem is solved when $V(a) \to \text{const}$ (may be
equal to zero) as $a \to 0$. If we find trajectories for which
$y = \dot{a}$ goes to infinity as $x = a \to 0$, then the horizon is present in
such a model. This is the case of dust brane model as well as
a brane world filled with domain-wall-like matter
with $\gamma = 1/3$.

In general our model does not solve the horizon problem due to
the existence of matter dominated phase of the early evolution where
the term $\rho^2$ dominates and so $H \propto \rho \propto
a^{-3\gamma}$ ($a(t) \propto t^{-1/3\gamma}$). Only in the
special case of heavy domain-wall-like matter ($\gamma \le 1/3$)
the horizon problem is solved, because the evolution near the singularity
dominated by matter (brane tension) is $\dot{a} \propto a^{1-3\gamma}$.
However, the existence of such a singularity requires ${\cal U} =0$ (Weyl tensor contribution
from the bulk vanishes). Therefore, the case $\gamma \le 1/3$ is strictly distinguished
because as $x \to 0$ then $y \to \text{const}$. For $\gamma = 1/3$ the term
$\rho^{2}/\lambda$ vanishes and brane effects are negligible.

In the phase portraits we can observe similarities as well as
differences in both considered cases. In the dust-like matter case
there is a quasi-static stage first discussed by Lema{\^\i}tre
(called `loitering stage' in \cite{Sahni92}). Such quasi-static
stages, present in the case of wall-like-matter, corresponds to
configurations in the vicinity of a critical point.
The acceleration does not depend on the value of $\Omega_{k}$, but
in the phase space there are different sets of initial conditions
which provide acceleration.

We conclude that the classes of admissible models in
brane scenario are {\it qualitatively} the same as in general
relativity, i.e., there exists a homeomorphism in the phase space
which transforms trajectories of brane models into general
relativistic ones without a change of orientation. However, {\it
quantitatively} the models differ from general relativistic ones -
for instance, they asymptote the singularity in a different way.

Finally, we reformulated the phase space quantities interms of the
observational parameters $\Omega$ and presented the formalism in
which one is able to perform the observational tests
(redshift-magnitude relation) of brane models which we will
address in a separate paper.

\section{Acknowledgements}

M.S. and A.K. were partially supported by the Polish Research
Committee (KBN) Grant No 2PO3B 107 22.

\end{document}